\documentclass[12pt]{iopart}
\eqnobysec
\usepackage{iopams,amsthm,graphicx,cite}

\begin{document}

\def\llra{\relbar\joinrel\longrightarrow}              
\def\mapright#1{\smash{\mathop{\llra}\limits_{#1}}}    
\def\mapup#1{\smash{\mathop{\llra}\limits^{#1}}}     
\def\mapupdown#1#2{\smash{\mathop{\llra}\limits^{#1}_{#2}}} 

\catcode`\@=11

\def\BF#1{{\bf {#1}}}
\def\NEG#1{{\rlap/#1}}

\def\Let@{\relax\iffalse{\fi\let\\=\cr\iffalse}\fi}
\def\vspace@{\def\vspace##1{\crcr\noalign{\vskip##1\relax}}}
\def\multilimits@{\bgroup\vspace@\Let@
 \baselineskip\fontdimen10 \scriptfont\tw@
 \advance\baselineskip\fontdimen12 \scriptfont\tw@
 \lineskip\thr@@\fontdimen8 \scriptfont\thr@@
 \lineskiplimit\lineskip
 \vbox\bgroup\ialign\bgroup\hfil$\m@th\scriptstyle{##}$\hfil\crcr}
\def\Sb{_\multilimits@}
\def\endSb{\crcr\egroup\egroup\egroup}
\def\Sp{^\multilimits@}
\let\endSp\endSb


\title[]{Reply to ``Comment on `On the inconsistency of the Bohm-Gadella
theory with quantum mechanics'\,''}

\author{Rafael de la Madrid}
\address{Department of Physics, University of California at San Diego,
La Jolla, CA 92093 \\
E-mail: \texttt{rafa@physics.ucsd.edu}}

\date{\small{(March 14, 2007)}}

\begin{abstract}
In this reply, we show that when we apply standard distribution theory
to the Lippmann-Schwinger equation, the resulting spaces of test functions
would comply with the Hardy axiom only if classic results of 
Paley and Wiener, of Gelfand and Shilov, and of the theory of 
ultradistributions were wrong. As well, we point out several differences 
between the ``standard method'' of constructing rigged Hilbert spaces in 
quantum mechanics and the method used in Time Asymmetric Quantum Theory.
\end{abstract}



\section{Introduction}
\label{sec:intro}

The authors of~\cite{C} allege to have shown that the conclusions 
of~\cite{HARDY} regarding the inconsistency of Time Asymmetric Quantum 
Theory (TAQT) with quantum mechanics are false. In this reply, we will
show that the arguments of~\cite{C} are missing essential aspects 
of~\cite{HARDY}, and that therefore the conclusions of~\cite{HARDY}
still stand.

The most important claims of~\cite{C} are the following:
\begin{enumerate}
     \item[{\bf 1}.] There are many examples of TAQT, and the present author 
has inadvertently constructed another one.
     \item[{\bf 2}.] The flaws of the Quantum Arrow of Time (QAT) 
pointed out in~\cite{HARDY} are actually not flaws, because the original 
derivation of the QAT was misquoted from its source~\cite{JMP95}.
      \item[{\bf 3}.] The crucial argument of~\cite{HARDY} regarding the
exponential blowup of the test functions $\widehat{\varphi}^{\pm}(z)$
does not prevent $\widehat{\varphi}^{\pm}(z)$ from being of Hardy class.
\end{enumerate}     
 
As we shall see, all these claims do not stand close scrutiny. In order to
show why, in Sec.~\ref{sec:stamet} we will outline the method to
construct rigged Hilbert spaces in quantum mechanics based on the
theory of distributions~\cite{GELFAND}. We shall refer to this method
as the ``standard method'' and show that the resulting rigged Hilbert spaces
are not of Hardy class. We shall also explain the
meaning of the exponential blowup of $\widehat{\varphi}^{\pm}(z)$ and why
it implies that the spaces of test functions are not of Hardy class. In
Sec.~\ref{sec:TAQTvsSQM}, we briefly outline the method to introduce
rigged Hilbert spaces of Hardy class in TAQT and compare such method with
the ``standard method.'' It will then be apparent that using the
method of TAQT, one can introduce any arbitrary rigged Hilbert space 
for the Gamow states. In order to address claim~{\bf 2}, we show (again) 
in Sec.~\ref{seec:QAT} that no matter how one introduces it, the Quantum
Arrow of Time has little to do with the actual time evolution of a quantum 
system. To address claim~{\bf 3}, in Sec.~\ref{sec:clasins} we use classic 
results of Paley and Wiener and of Gelfand and Shilov to show that the 
``standard method'' of dealing with the Lippmann-Schwinger equation leads to 
rigged Hilbert spaces that are {\it not} of Hardy class. Section~\ref{sec:con} 
concludes that the arguments of~\cite{HARDY} still stand.

\section{The ``standard method''}
\label{sec:stamet}

In this section, we illustrate the main features of the ``standard method''
to construct rigged Hilbert spaces in quantum 
mechanics~\cite{RELBO}. Such ``standard
method'' is based on the theory of distributions~\cite{GELFAND}. For the sake 
of clarity, we shall use the spherical shell potential of height $V_0$,
\begin{equation}
	V(\vec{x})=V(r)=\left\{ \begin{array}{ll}
                                0   &0<r<a  \\
                                V_0 &a<r<b  \\
                                0   &b<r<\infty \, .
                  \end{array} 
                 \right. 
	\label{potential}
\end{equation}
For $l=0$, the Hamiltonian acts as (we take $\hbar ^2/2m=1$)
\begin{equation}
      H = -\frac{\rmd ^2}{\rmd r^2} + V(r)    \, .  
  \label{doh}
\end{equation}
The regular solution is
\begin{equation}
      \chi (r;E)=\left\{ \begin{array}{lll}
               \sin (\sqrt{E \,} r) \quad &0<r<a  \\
               {\cal J}_1(E)\rme ^{\rmi \sqrt{E-V_0 \,} r}
                +{\cal J}_2(E)\rme ^{-\rmi \sqrt{E-V_0 \,}  r}
                 \quad  &a<r<b \\
               {\cal J}_3(E) \rme ^{\rmi \sqrt{E \,} r}
                +{\cal J}_4(E)\rme ^{-\rmi \sqrt{E \,}  r}
                 \quad  &b<r<\infty \, .
               \end{array} 
                 \right. 
             \label{chi}
\end{equation}
The Jost functions and the $S$ matrix are given by
\begin{equation}
      {\cal J}_+(E)=-2\rmi {\cal J}_4(E) \, , \quad 
      {\cal J}_-(E)=2\rmi {\cal J}_3(E) \, ,
        \label{josfuc1}
\end{equation}
\begin{equation}
       S(E) =\frac{{\cal J}_-(E)}{{\cal J}_+(E)} \, .
       \label{smatrix1}
\end{equation}
The solutions of the Lippmann-Schwinger equation can be written as
\begin{equation}
      \langle r|E^{\pm}\rangle \equiv 
       \chi ^{\pm}(r;E)= \sqrt{\frac{1}{\pi}
         \frac{1}{\sqrt{E \,}\,}\,}  \,  
     \frac{\chi (r;E)}{{\cal J}_{\pm}(E)}  \, .
      \label{LSdnesb}
\end{equation}
When $V$ tends to zero, these eigensolutions tend to the ``free'' 
eigensolution:
\begin{equation}
      \langle r|E\rangle  \equiv 
       \chi _0(r;E)= \sqrt{\frac{1}{\pi}
         \frac{1}{\sqrt{E \,}\,}\,} \, \sin (\sqrt{E\,}r)  \, .
      \label{LSdnesb0}
\end{equation}
These eigenfunctions are delta-normalized and therefore their associated 
unitary operators,
\begin{equation}
      (U_{\pm}f)(E)=\int_0^{\infty}\rmd r \, 
                         \overline{\chi ^{\pm}(r;E)}\, f(r) \equiv
          \widehat{f}_{\pm}(E) \, , \quad E\geq 0 \, ,  
         \label{Upm}
\end{equation}
\begin{equation}
      (U_0f)(E)=\int_0^{\infty}\rmd r \, 
                         \overline{\chi _0(r;E)}\, f(r) \equiv
          \widehat{f}_0(E) \, , \quad E\geq 0 \, ,  
         \label{Upm0}
\end{equation}
transform from $L^2([0,\infty),\rmd r)$ onto $L^2([0,\infty),\rmd E)$. 

The Lippmann-Schwinger and the ``free'' eigenfunctions can be analytically 
continued from the scattering spectrum into the whole complex plane. We shall 
denote such analytically continued eigenfunctions by $\chi ^{\pm}(r;z)$
and $\chi _0(r;z)$. Whenever they exist, the analytic continuations 
of~(\ref{Upm}) and (\ref{Upm0}) are denoted by
\begin{equation}
      \widehat{f}_{\pm}(z)=\int_0^{\infty}\rmd r \, 
                         \overline{\chi ^{\pm}(r;\overline{z})}\, f(r)  \, ,  
         \label{Upmac}
\end{equation}
\begin{equation}
      \widehat{f}_{0}(z)=\int_0^{\infty}\rmd r \, 
                         \overline{\chi _0(r;\overline{z})}\, f(r)  \, ,  
         \label{Upm0ac}
\end{equation}
where here and in the following $z$ belongs to a two-sheeted Riemann surface.

The resonant energies are given by the poles $z_n$ of the $S$ matrix, and their
associated Gamow states are
\begin{equation}
	u(r;z_n) = N_n\left\{ \begin{array}{ll}
         \frac{1}{{\mathcal J}_3(z_n)}\sin(\sqrt{z_n\,}r)  &0<r<a \\ [1ex]
         \frac{{\mathcal J}_1(z_n)}{{\mathcal J}_3(z_n)}
            \rme ^{\rmi \sqrt{z_n-V_0\,}r}
         +\frac{{\mathcal J}_2(z_n)}{{\mathcal J}_3(z_n)}
                \rme ^{-\rmi \sqrt{z_n-V_0\,}r} 
                  &a<r<b 
         \\  [1ex]
         \rme ^{\rmi \sqrt{z_n\,}r}  &b<r<\infty \, ,
                           \end{array} 
                  \right.
	\label{dgv0p} 
\end{equation}
where $N_n$ is a normalization factor.

The theory of distributions~\cite{GELFAND} says that a test function 
$\varphi (r)$ on which 
a distribution $d(r)$ acts is such that the following
integral is finite:\footnote{In quantum mechanics, we need to impose a few
more requirements, but we will not need to go into such details here.}
\begin{equation}
     \langle \varphi|d\rangle \equiv 
        \int_0^{\infty}\rmd r \, \overline{\varphi (r)} d(r) <\infty \, ,
          \label{basic}
\end{equation}
where $\langle \varphi |d\rangle$ represents the action of the
functional $|d\rangle$ on the test function $\varphi$. With some variations, 
this is the ``standard method'' followed 
by~\cite{SUDARSHAN,BOLLINI,FP02,JPA02,IJTP03,JPA04,EJP05,LS1,LS2} to introduce 
spaces of test functions in quantum mechanics. Thus, contrary to what the
authors of~\cite{C} assert, the method followed by the present author runs
(somewhat) parallel to~\cite{BOLLINI}, not to~TAQT.

In order to use~(\ref{basic}) to construct the rigged Hilbert 
spaces for the analytically continued Lippmann-Schwinger eigenfunctions and
for the Gamow states, we need to obtain the growth of $\chi ^{\pm}(r;z)$, 
$\chi _0(r;z)$ and $u(r;z_n)$. Because the regular solution blows up 
exponentially~\cite{TAYLOR}, 
\begin{equation}
      \left| \chi (r; z)\right| \leq C \, 
        \frac{\left|z\right|^{1/2}r}{1+\left|z\right|^{1/2}r} \,  
      \rme ^{|{\rm Im}\sqrt{z\,}|r}  \, ,   
      \label{boundrs}
\end{equation}
the growth of the eigenfunctions~(\ref{LSdnesb}), (\ref{LSdnesb0}) and 
(\ref{dgv0p}) blows up exponentially:
\begin{equation}
      |\chi ^{\pm}(r;z)| \leq  C \, \frac{1}{{\cal J}_{\pm}(z)} \, 
        \frac{\left|z\right|^{1/4}r}{1+\left|z\right|^{1/2}r} \,  
      \rme ^{|{\rm Im}\sqrt{z\,}|r}   \, , 
          \label{boundpms}  
\end{equation}
\begin{equation}
      |\chi _0(r;z)|  \leq  C  \, 
        \frac{\left|z\right|^{1/4}r}{1+\left|z\right|^{1/2}r} \,  
      \rme ^{|{\rm Im}\sqrt{z\,}|r}   \, ,   
\end{equation}
\begin{equation}
      |u(r;z_n)| \leq C_n \, 
           \frac{\left|z _n\right|^{1/2}r}{1+\left|z_n \right|^{1/2}r}
                \rme ^{|{\rm Im}\sqrt{z_n\,}|r} \, .
\end{equation}
When we plug this exponential blowup into the basic requirement~(\ref{basic}) 
of the ``standard method,'' we see that the test functions on which those
distributions act must fall off at least exponentially. 

By using the Gelfand-Shilov theory of $M$ an $\Omega$ 
functions~\cite{GELFAND}, it was shown in~\cite{LS2} that when $a$ and $b$ 
are positive real numbers satisfying
\begin{equation}
       \frac{1}{a}+\frac{1}{b} = 1 \, ,
           \label{ab}
\end{equation}
and when $\varphi ^+(r)$ is an infinitely differentiable function
whose tails fall off like $\rme ^{-r^a/a}$, then $\varphi ^+(z)$ grows
like $\rme ^{|{\rm Im}(\sqrt{z})|^b/b}$ in the infinite arc of the lower 
half-plane of the Riemann surface:
\begin{equation}
      \hskip-1cm {\rm If} \ |\varphi ^+(r)| < C \rme ^{-\frac{\, r^a}{a}}
        \  {\rm as} \ r \to \infty , \ {\rm then} \
      |\widehat{\varphi}^+(z)| \leq
            C \rme ^{\frac{\, |{\rm Im}(\sqrt{z})|^b}{b}} \
       {\rm as} \ |z| \to \infty  \,.
        \label{blowupfex}
\end{equation}
It was shown in~\cite{HARDY} that when $\varphi ^+(r) \in C_0^{\infty}$, 
$\widehat{\varphi}^+(z)$ blows up exponentially in the infinite arc of 
the lower half-plane of the Riemann surface:
\begin{equation} 
     {\rm If} \  |\varphi ^+(r)| = 0 \ {\rm when} \ r>A , \ 
      {\rm then} \ |\widehat{\varphi}^+(z)| \leq C 
               \rme ^{A|{\rm Im}\sqrt{z\,}|} 
       \ {\rm as} \ |z| \to \infty  \, .
        \label{grofainf}
\end{equation}
From the above estimates, we concluded in~\cite{HARDY} 
that the $\varphi ^+$'s obtained from the ``standard method'' cannot be 
Hardy functions, since $\widehat{\varphi}^+(z)$ does not tend to zero as 
$|z|$ tends to infinity. 

The authors of~\cite{C} argue that one cannot draw any conclusion 
on the limit $|z|\to \infty$ from estimates such as~(\ref{blowupfex}) or 
(\ref{grofainf}), and therefore they conclude that nothing
prevents $\widehat{\varphi}^+(z)$ from tending to zero and therefore from
being Hardy functions. Their conclusion is not true, because their argument 
does not take the nature of~(\ref{blowupfex}) and 
(\ref{grofainf}) into account. After we explain the meaning of those estimates,
it will be clear why they prevent $\widehat{\varphi}^{\pm}(z)$ from tending 
to zero in any infinite arc of the Riemann surface.

In order to understand what~(\ref{blowupfex}) and 
(\ref{grofainf}) mean, we start with the simple sine function 
$\sin (\sqrt{E\, }r)$. When $E\geq 0$, the sine function oscillates between 
$+1$ and $-1$:
\begin{equation}
      |\sin (\sqrt{E\, }r)| \leq 1 \, , \quad E\geq 0 \, .
\end{equation}
As $E$ tends to infinity, such oscillatory behavior remains, and in such limit
the sine function does not tend to zero. When we analytically 
continue the sine function,
\begin{equation}
      \sin (\sqrt{z\, }r) \, , 
            \label{sineac}
\end{equation}
the oscillations are bounded by
\begin{equation}
      |\sin (\sqrt{z\, }r)| \leq  C \, 
        \frac{\left|z\right|^{1/2}r}{1+\left|z\right|^{1/2}r} \,  
      \rme ^{|{\rm Im}\sqrt{z\,}|r}  \, .
            \label{sineabv}
\end{equation}
Thus, as $|z|$ tends to infinity, $\sin (\sqrt{z\,}r)$ oscillates wildly, and
the magnitude of its oscillation is tightly bounded by the exponential 
function. It is certain that as $|z|$ tends to infinity, $\sin (\sqrt{z\,}r)$ 
does not tend to zero, even though the function vanishes when 
$\sqrt{z\, }r = \pm n \pi$, $n=0,1,\ldots$

It just happens that the solutions of the Lippmann-Schwinger equation follow
the same pattern. When $E$ is positive, the eigensolutions are oscillatory
and bounded by
\begin{equation}
      |\chi ^{\pm}(r;E)| \leq  C \, \frac{1}{{\cal J}_{\pm}(E)} \, 
        \frac{\left|E\right|^{1/4}r}{1+\left|E\right|^{1/2}r}    \, .   
\end{equation}
When the energy is complex, their oscillations get wild and are bounded 
by Eq.~(\ref{boundpms}).\footnote{The points at which
${\cal J}_{\pm}(z)=0$ do not affect the essence of the argument.} Thus, the 
analytic continuations of the Lippmann-Schwinger eigenfunctions
oscillate wildly, and the magnitude of their oscillation is tightly bounded
by an exponential function (multiplied by factors that don't cancel the
exponential blowup when $|z|\to \infty$).

Because in Eqs.~(\ref{Upmac}) and (\ref{Upm0ac}) we are integrating
over $r$, the exponentially-bounded oscillations of 
$\chi ^{\pm}(r;z)$ get transmitted into $\widehat{\varphi}^{\pm}(z)$. The 
estimates~(\ref{blowupfex}) and (\ref{grofainf}) bound the oscillation
of the test functions of the ``standard method,'' except for
factors that don't cancel the exponential blowup. It is the 
exponentially-bounded oscillations of $\widehat{\varphi}^{\pm}(z)$ what
prevent $\widehat{\varphi}^{\pm}(z)$ from tending to
zero in any infinite arc of the Riemann surface and therefore from being
of Hardy class.

A somewhat simpler way to understand the above estimates is by looking at
the ``free'' incoming and outgoing wave functions 
${\varphi}^{\rm in}$ and $\varphi ^{\rm out}$. Because
in the energy representation such wave functions are the same as the
``in'' and ``out'' wave functions, 
\begin{equation}
      \widehat{\varphi}^{\rm in}(E) = \langle E|{\varphi}^{\rm in}\rangle
   =\langle ^+E|{\varphi}^{+}\rangle = \widehat{\varphi}^{+}(E)\, , 
      \label{asstrco1}
\end{equation}
\begin{equation}
    \widehat{\varphi}^{\rm out}(E) = \langle E|{\varphi}^{\rm out}\rangle
   =\langle ^-E|{\varphi}^{-}\rangle =\widehat{\varphi}^{-}(E)  \, ,
       \label{asstrco2}
\end{equation}
in TAQT the analytic continuation of
$\widehat{\varphi}^{\rm in}(E)$ and $\widehat{\varphi}^{\rm out}(E)$ are also
of Hardy class. Since 
\begin{equation}
     \widehat{\varphi}^{\rm in, out}(z) =\int_0^{\infty}\rmd r \, 
          \frac{1}{\sqrt{\pi}} \frac{1}{z^{1/4}} \, 
                     \sin (\sqrt{z}\,r)  \,\varphi ^{\rm in, out}(r)  \, ,  
         \label{Upm0in}
\end{equation}
it is evident that the exponential blowup~(\ref{sineabv}) of
$\sin (\sqrt{z\,}r)$ will prevent $\widehat{\varphi}^{\rm in, out}(z)$ from
tending to zero as $|z| \to \infty$ in any half-plane of the Riemann 
surface. Thus, $\widehat{\varphi}^{\rm in, out}(z)$ are not of Hardy class,
contrary to TAQT.

Strictly speaking, the bounds~(\ref{blowupfex}) and (\ref{grofainf}) are
not the tightest ones. We should include polynomial corrections, see 
Eq.~(B.15) in~\cite{LS2}, and the effect of
$\frac{\left|z\right|^{1/4}r}{1+\left|z\right|^{1/2}r}$ and 
$\frac{1}{{\cal J}_{\pm}(z)}$ to obtain the tightest bounds. We shall not
obtain those corrections here, because they do not cancel
the exponential blowup at infinity, and because in this reply we shall
use instead other classic bounds, see Sec.~\ref{sec:clasins}.

Let us summarize this section. In standard quantum mechanics, once the 
Lippmann-Schwinger equation is solved, the properties of 
$\widehat{\varphi}^{\pm}(z)$ are already determined by 
Eqs.~(\ref{Upmac}) and (\ref{Upm0ac}), and there 
is no room for any extra assumption on their properties. This means, in 
particular, that the Hardy axiom cannot be 
simply assumed. Rather, the Hardy axiom must be proved using
Eqs.~(\ref{Upmac}) and (\ref{Upm0ac}).\footnote{This is what in~\cite{HARDY} 
it was meant by the assertion that the Hardy axiom is not a matter of 
assumption but a matter of proof.} It simply happens that the 
``standard method'' yields $\widehat{\varphi}^{\pm}(z)$
and $\widehat{\varphi}^{\rm in,out}(z)$ that oscillate wildly. Because these 
oscillations are bounded by exponential functions, $\widehat{\varphi}^{\pm}(z)$
and $\widehat{\varphi}^{\rm in,out}(z)$ do not tend to zero as $|z|$ tends 
to infinity in any half-plane of the Riemann surface---hence they are not of 
Hardy class.

\section{TAQT vs.~the ``standard method''}
\label{sec:TAQTvsSQM}

In TAQT, one doesn't solve the Lippmann-Schwinger equation in order to 
afterward obtain the properties of $\widehat{\varphi}^{\pm}(z)$ 
using Eq.~(\ref{Upmac}). Instead, one transforms into the
energy representation (using $U_{\pm}$ in our example) and then imposes
the Hardy axiom. If ${\cal H}_{\pm}^2$ denotes the
spaces of Hardy functions from above ($+$) and below ($-$), $\cal S$ 
denotes the Schwartz space, and $\tilde{\Phi}_{\pm}$ denote their intersection
restricted to the positive real line,
\begin{equation}
      \tilde{\Phi}_{\pm} = {\cal H}_{\pm}^2\cap {\cal S}|_{{\mathbb R}^+} \, ,
\end{equation}
then the Hardy axiom states that the functions $\widehat{\varphi}^{\pm}(z)$
belong to $\tilde{\Phi}_{\mp}$:
\begin{equation}
       \widehat{\varphi}^{\pm}(z) \in \tilde{\Phi}_{\mp} \, .
          \label{sssusm}
\end{equation}
This means that in the position representation, the Gamow
states and the analytic continuation of the Lippmann-Schwinger eigenfunctions 
act on the following spaces:
\begin{equation}
       \Phi _{{\rm BG}\mp}= U_{\pm}^{-1} \tilde{\Phi}_{\mp}  \, .  
     \label{BGchoice}
\end{equation}
It is obvious that the choices~(\ref{sssusm})-(\ref{BGchoice}) are 
arbitrary. One may as well choose another dense subset of 
$L^2([0,\infty ),\rmd E)$ with different properties
and obtain a different space of test functions for the Gamow states. What is 
more, $\Phi _{{\rm BG}\pm}$ are different from the spaces of test
functions obtained through the ``standard method,'' because the
functions $\widehat{\varphi}^{\pm}(z)$ of the ``standard method'' are not 
of Hardy class.

The authors of~\cite{C} claim that the present author has inadvertently 
constructed an example of TAQT. That such is not the case can be seen not 
only from the differences between the ``standard method'' and the method used 
in TAQT to introduce rigged Hilbert spaces, but also from the
outcomes. For example, whereas in the position representation the 
``standard method'' calls for just {\it one} rigged Hilbert space for the Gamow
states and for the analytically continued Lippmann-Schwinger 
eigenfunctions~\cite{LS2}, TAQT uses {\it two} rigged Hilbert spaces
\begin{equation}
        \Phi _{{\rm BG}\pm} 
          \subset L^2([0,\infty ),\rmd r) \subset
         \Phi _{{\rm BG}\pm}^{\times} \,.
     \label{BGchoicetwo}
\end{equation}
One of the rigged Hilbert spaces is used for the ``in'' solutions and for
the anti-resonant states, whereas the other one is used for the ``out'' 
solutions and for the resonant states. Another difference is that in TAQT, the 
solutions of the 
Lippmann-Schwinger equation for scattering energies have a time asymmetric 
evolution~\cite{BLUNDER}, whereas the ``standard method'' yields that
such time evolution runs from $t=-\infty$ to $t=+\infty$, 
see~\cite{LS1}. Incidentally, this is an instance where TAQT differs not only 
mathematically but also physically from standard quantum mechanics, because in 
standard scattering theory, the time evolution of a scattering process goes 
from the asymptotically remote past ($t \to -\infty$) to the asymptotically far
future ($t \to +\infty$). This is not so in TAQT~\cite{BLUNDER}.

It seems hardly necessary to clarify what the present author means by 
``standard quantum mechanics.'' Standard quantum mechanics means the 
Schr\"odinger
equation, and standard scattering theory means the Lippmann-Schwinger
equation. In standard quantum mechanics, one assumes that these equations 
describe the physics and then solves them. Because of the scattering and 
resonant spectra, their solutions lie within rigged Hilbert spaces. The 
construction of such rigged Hilbert spaces follows by application of the
``standard method.'' By contrast, TAQT simply assumes that the solutions of 
the Schr\"odinger and the Lippmann-Schwinger equations comply with the Hardy 
axiom, without ever showing that the actual solutions of those equations
comply with such axiom.

It was claimed in~\cite{HARDY} that there is no example of TAQT. The
authors of~\cite{C} dispute such claim and assert that there are
many examples. The present author disagrees with their assertion, because
{\it assuming} that for a large class of potentials the solutions of the 
Lippmann-Schwinger equation comply with the Hardy axiom is not the same as
having an example where it is shown that the {\it actual} solutions of the
Lippmann-Schwinger equation comply with the Hardy axiom. In fact, 
to the best of the present author's knowledge, no advocate of TAQT has ever
used Eq.~(\ref{Upmac}) to discuss the analytic properties of 
$\widehat{\varphi}^{\pm}(E) = \langle ^{\pm}E|\varphi ^{\pm}\rangle$ in 
terms of the actual solutions $\chi^{\pm}(r;E)$ of the Lippmann-Schwinger 
equation. 

The authors of~\cite{C} inadvertently acknowledge that there is no example 
of TAQT when they say that they still need ``{\it to identify the
form and properties}'' of the functions of~(\ref{BGchoice}), see the last
paragraph in section~2 of~\cite{C}. By saying so, they are
acknowledging that they don't know whether the standard Gamow states defined
in the position representation are well defined as functionals acting
on $\Phi _{{\rm BG}\pm}$. If TAQT had an example, it would be known.

\section{The Quantum Arrow of Time (QAT)}
\label{seec:QAT}

Advocates of TAQT argue that their choice~(\ref{BGchoice}) is not
arbitrary but rather is rooted on a causality principle. Such causality 
principle is the ``preparation-registration arrow of time,'' sometimes referred
to as the ``Quantum Arrow of Time'' (QAT). For the ``in'' states
$\varphi ^+$, the causal statement of the QAT is written as
\begin{equation}
     \widetilde{\varphi}^+(t) \equiv \int_{-\infty}^{+\infty}\rmd E \,   
       \rme ^{-\rmi Et} \widehat{\varphi}^{+}(E) =0
            \, , \quad \mbox{for} \ t>0  \, .
      \label{ferFtvph+}
\end{equation}
By one of the Paley-Wiener theorems, Eq.~(\ref{ferFtvph+}) is equivalent to
assuming that $\widehat{\varphi}^+(E)$ is of Hardy class from below. The
corresponding causal statement for the ``out'' wave functions $\varphi ^-$ 
implies that $\varphi ^-$ is of Hardy class from above. Hence, in TAQT, the 
choice~(\ref{BGchoice}) is not arbitrary but a consequence of causality.

It was pointed out in~\cite{HARDY} that the QAT is flawed. The argument was
twofold. First, it was pointed out that the original derivation~\cite{JMP95}
of Eq.~(\ref{ferFtvph+}) made use of the following flawed assumption:
\begin{equation}
     0=\langle E|\varphi ^{\rm in}(t)\rangle =
     \langle ^+E|\varphi ^{+}(t)\rangle = 
            \rme ^{-\rmi Et} \widehat{\varphi}^{+}(E)
     \, , \quad {\rm for \ all \ energies,} 
        \label{flawassum1} 
\end{equation}
which can happen only when $\varphi ^+$ and $\varphi ^{\rm in}$ are 
identically 0. It was then pointed out that even though one may simply assume 
the causal statement~(\ref{ferFtvph+}) and forget about how it was derived, 
such causal statement says little about the actual time evolution of a quantum
system, because the quantum mechanical time evolution of $\varphi ^+$ is not
given by Eq.~(\ref{ferFtvph+}):
\begin{equation}
      \varphi ^+(t)= \rme ^{-\rmi Ht} \varphi ^+ \ \neq \   
       \widetilde{\varphi}^+(t)  \, .
         \label{neoe}
\end{equation}

To counter this argument, the authors of~\cite{C} claim that the derivation 
of the QAT was misquoted from the original source~\cite{JMP95}, and that the 
flawed assumption~(\ref{flawassum1}) was never used to derive
the QAT~(\ref{ferFtvph+}). It seems therefore necessary to quote the original
derivation (see~\cite{JMP95}, page~2597):\footnote{In this quote, 
$\phi ^{\rm in}$, $\phi ^+$, ${\cal F}(t)$ and Eq.~(\ref{Ftdkd}) correspond, 
respectively, to $\varphi ^{\rm in}$, $\varphi ^+$, $\widetilde{\varphi}^+(t)$ 
and Eq.~(\ref{ferFtvph+}).}
\begin{quote}
{\small \it ``We are now in the position to give a mathematical formulation of 
the QAT: we choose $t=0$ to be the time before which all preparations of
$\phi ^{\rm in}(t)$ are completed and after which the registration of
$\psi ^{\rm out}(t)$ begins. This means that for $t>0$ the energy distribution
of the preparation apparatus must vanish: 
$\langle E,\eta |\phi ^{\rm in}(t) \rangle =0$ for all values of the quantum
numbers $E$ and $\eta$ ($\eta$ are the additional quantum numbers which we
usually suppress). As the mathematical statement for `no preparations for
$t>0$' we therefore write (the slightly weaker condition)
\begin{equation}
     \hskip-0.5cm  0= \int \rmd E \, \langle E|\phi ^{\rm in}(t)\rangle =
       \int \rmd E \, \langle ^+E|\phi ^{+}(t)\rangle =
       \int \rmd E \, \langle ^+E|\rme ^{-\rmi Ht}|\phi ^{+}\rangle  
\end{equation}
or
\begin{equation}
     \hskip-0.5cm 
       0= \int_{-\infty}^{+\infty} \rmd E \, 
              \langle ^+E|\phi ^{+}\rangle \rme ^{-\rmi Et} 
          \equiv {\cal F}(t) \quad {\rm for \ } t>0  \, .
         \label{Ftdkd}
\end{equation}
   }
\end{quote}
The readers can decide whether or not the flawed 
hypothesis~(\ref{flawassum1}) was used to derive the QAT~(\ref{Ftdkd}).

Nevertheless, it is actually not very relevant whether the authors 
of~\cite{JMP95} used~(\ref{flawassum1}) to derive~(\ref{ferFtvph+}). As 
pointed out in~\cite{HARDY}, and as mentioned 
above, even though one can forget~(\ref{flawassum1}) and simply 
assume~(\ref{ferFtvph+}) as the causal condition to be satisfied by 
$\varphi ^+$, such causal condition has little to do with the time evolution 
of a quantum system, see again Eq.~(\ref{neoe}). In particular, as even the
author of~\cite{BAUMGARTEL} has asserted, the $t$ that appears in 
Eq.~(\ref{ferFtvph+}) is not the same as the parametric time $t$
that labels the evolution of a quantum system.\footnote{All this shows that 
the new term TAQT is a misnomer. A better name is Bohm-Gadella theory, because 
it was these two authors who proposed the theory and summarized it 
in~\cite{BG}.} Thus, as far as standard quantum mechanics is concerned, the 
causal content of the QAT is physically vacuous, and therefore, regardless of 
how one motivates it, there is no physical justification for the 
choice~(\ref{BGchoice}).

\section{TAQT vs.~the ``classic results''}
\label{sec:clasins}

In this section, we are going to compare the Hardy axiom of TAQT with
some classic results of Paley and Wiener, of Gelfand and Shilov and of
the theory of ultradistributions, which we shall collectively refer to
as the ``classic results.'' More precisely, we will see that the spaces of
test functions $\widehat{\varphi}^{\pm}$ obtained by the ``standard method''
would be of Hardy class only if the ``classic results'' were wrong.

The direct comparison with the ``classic results'' is more easily done
in one dimension, and therefore we shall use the example of the one-dimensional
rectangular barrier potential:
\begin{equation}
           V(x)=\left\{ \begin{array}{ll}
                                0   &-\infty <x<a  \\
                                V_0 &a<x<b  \\
                                0   &b<x<\infty \, .
                  \end{array} 
                 \right. 
	\label{sbpotential1}
\end{equation}
For this potential, the ``in'' and ``out'' eigensolutions are well known
and can be found for example in~\cite{JPA04}. We shall denote them by
$\chi _{\rm l,r}^{\pm}(x;E)$, where the labels l,r denote left and right 
incidence. When we analytically continue these eigenfunctions, or when
we consider the Gamow states for this potential, the ``standard method''
calls for test functions $\varphi _{\rm l,r}^{\pm}(x)$ for which the 
following integrals are finite:
\begin{equation}
      \widehat{\varphi}_{\rm l,r}^{\pm}(z)=\int_{-\infty}^{\infty}\rmd x \, 
           \overline{\chi _{\rm l,r}^{\pm}(x;\overline{z})}\, {\varphi}(x)  
                  \, .
         \label{Upmac1D}
\end{equation}
Just as in the example discussed in Sec.~\ref{sec:stamet}, the
test functions $\varphi (x)$ must at least fall off
faster than exponentials.

To further simplify the discussion, we need to recall that, because
of Eqs.~(\ref{asstrco1}) and (\ref{asstrco2}), the Hardy axiom assumes
that the ``free'' wave functions $\widehat{\varphi}_{\rm l,r}^{\rm in}(E)$ and 
$\widehat{\varphi}_{\rm l,r}^{\rm out}(E)$ 
are also of Hardy class. These ``free'' functions are given by (hereafter, we
just consider $\varphi _{\rm l,r}^{\rm in}$, since the analysis
for $\varphi _{\rm l,r}^{\rm out}$ is the same)
\begin{equation}
      \widehat{\varphi}_{\rm l}^{\rm in}(E)=
                    \frac{1}{\sqrt{4\pi k \,}}
                  \int_{-\infty}^{\infty}\rmd x \,
                   \rme ^{-\rmi kx} \, {\varphi}^{\rm in}(x)  
                  \, ,
         \label{Upmac1D0in}
\end{equation}
\begin{equation}
      \widehat{\varphi}_{\rm r}^{\rm in}(E)=
               \frac{1}{\sqrt{4\pi k \,}} 
          \int_{-\infty}^{\infty}\rmd x \,
                   \rme ^{\rmi kx} \, {\varphi}^{\rm in}(x)  
                  \, ,
         \label{Upmac1D0inr}
\end{equation}
where $k=\sqrt{E}$ is the wave number. The total wave function is given 
by the sum of left and right components:
\begin{equation}
      \widehat{\varphi}^{\rm in}(E)= \widehat{\varphi}_{\rm l}^{\rm in}(E) +
                      \widehat{\varphi}_{\rm r}^{\rm in}(E) \, .
         \label{lrimoc}
\end{equation}
It is simpler to work with $k$ rather than with $E$ and define
\begin{equation}
      \widehat{\varphi}_{\rm l,r}^{\rm in}(k) \equiv
          \sqrt{2k \,} \,
            \widehat{\varphi}_{\rm l,r}^{\rm in}(E)
                  \, ;
         \label{Upmac1D0in2}
\end{equation}
that is,
\begin{equation}
      \widehat{\varphi}_{\rm l}^{\rm in}(k)= \frac{1}{\sqrt{2\pi}}
                  \int_{-\infty}^{\infty}\rmd x \,
                   \rme ^{-\rmi kx} \, {\varphi}^{\rm in}(x)  
                  \, , \quad k\geq 0 \, ,
         \label{Upmac1D0ink}
\end{equation}
\begin{equation}
      \widehat{\varphi}_{\rm r}^{\rm in}(k)= \frac{1}{\sqrt{2\pi}}
                  \int_{-\infty}^{\infty}\rmd x \,
                   \rme ^{\rmi kx} \, {\varphi}^{\rm in}(x)  
                  \, ,  \quad k\geq 0 \, .
         \label{Upmac1D0irdk}
\end{equation}
The ``total'' wave function in the wave-number representation,
$\widehat{\varphi}^{\rm in}(k)= \widehat{\varphi}_{\rm l}^{\rm in}(k) +
                      \widehat{\varphi}_{\rm r}^{\rm in}(k)$, is
thus the Fourier transform of $\varphi (x)$,
\begin{equation}
      \widehat{\varphi}^{\rm in}(k)= \frac{1}{\sqrt{2\pi}}
                  \int_{-\infty}^{\infty}\rmd x \,
                   \rme ^{-\rmi kx} \, {\varphi}^{\rm in}(x)  
                  \, , \quad k \in {\mathbb R} \, .
         \label{lrimockks}
\end{equation}
Its analytic continuation will be denoted as
\begin{equation}
      \widehat{\varphi}^{\rm in}(q)= \frac{1}{\sqrt{2\pi}}
                  \int_{-\infty}^{\infty}\rmd x \,
                   \rme ^{-\rmi qx} \, {\varphi}^{\rm in}(x)  
                  \, , \quad q \in {\mathbb C} \, .
         \label{Upmac1D0inlrqFT}
\end{equation}

At this point, we are ready to introduce two classic theorems. The first one
is due to Paley and Wiener (see Theorem IX.11 in~\cite{SIMON}):

\vskip0.5cm

\theoremstyle{plain}
\newtheorem*{Th1}{Theorem~1 (Paley-Wiener)}
\begin{Th1} 
An entire analytic function $\widehat{\varphi}(q)$ is the Fourier transform 
of a $C_0^{\infty}({\mathbb R})$ function $\varphi (x)$ with support in 
the segment $\{ x \, | \ |x|<A \}$
if, and only if, for each $N$ there is a $C_N$ so that
\begin{equation}
      |\widehat{\varphi}(q)|\leq 
           \frac{C_N \, \rme ^{A|{\rm Im}(q)|}}{ (1+|q|)^N} 
         \label{PWbound}
\end{equation}
for all $q \in {\mathbb C}$.
\end{Th1}

\vskip0.5cm

This theorem says that the Fourier transform of a $C_0^{\infty}$ function 
is an analytic function that grows exponentially,
and that such exponential growth is mildly corrected (but not canceled) 
by a polynomial falloff.

The second theorem we shall use is due to Gelfand and 
Shilov~\cite{GELFAND}. Before stating it, we need some definitions. Let 
$a$ and $b$ denote two positive real numbers satisfying~(\ref{ab}). Let 
us define $\Phi _{a,b}$ as the set of all differentiable
functions $\varphi (x)$ ($-\infty <x <\infty$) satisfying the inequalities
\begin{equation}
      \left| \frac{\rmd ^n \varphi (x)}{\rmd x ^n} \right| 
            \leq C_n \rme ^{-\alpha \frac{|x|^a}{a}}
\end{equation}
with constants $C_n$ and $\alpha >0$ which may depend on the function 
$\varphi$. Let us define the space $\widehat{\Phi}_{a,b}$ as the set of entire
analytic functions $\widehat{\varphi}(q)$, $q={\rm Re}(q)+\rmi \,{\rm Im}(q)$, 
which satisfy the inequalities
\begin{equation}
       |q^n\widehat{\varphi}(q)|\leq C_n 
        \rme ^{+\beta \frac{|{\rm Im}(q)|^b}{b}} 
            \, , 
       \label{GSbound}
\end{equation}
where the constants $C_n$ and $\beta >0$ depend on the function $\varphi$. It 
is obvious that the elements of $\Phi _{a,b}$ are functions that, together with
their derivatives, decrease at infinity faster than 
$\rme ^{-\frac{|x|^a}{a}}$, whereas the elements of 
$\widehat{\Phi}_{a,b}$ are analytic 
functions that grow exponentially at infinity as 
$\rme ^{+\frac{|{\rm Im}(q)|^b}{b}}$, except for a polynomial correction that
doesn't cancel the exponential blowup.

\vskip0.5cm

\theoremstyle{plain}
\newtheorem*{Th2}{Theorem~2 (Gelfand-Shilov)}
\begin{Th2} 
The space $\widehat{\Phi}_{a,b}$ is the Fourier transform of 
$\Phi _{a,b}$.
\end{Th2}

\vskip0.5cm

This theorem means that the smooth functions that fall off at infinity
faster than $\rme ^{-|x|^a/a}$ are, in Fourier space, analytic 
functions that grow exponentially like $\rme ^{+|{\rm Im}(q)|^b/b}$.

The bounds~(\ref{PWbound}) and (\ref{GSbound}) are to be understood in
the same way as the bounds~(\ref{blowupfex}) and (\ref{grofainf}). That is,
the bounds~(\ref{PWbound}) and (\ref{GSbound}) mean that 
$\widehat{\varphi}(q)$ is an oscillatory function
that grows exponentially in the infinite arc of the $q$-plane, the oscillation
being tightly bounded by Eqs.~(\ref{PWbound}) and (\ref{GSbound}) when 
$\varphi (x)$ belongs to $C_0^{\infty}$ and $\Phi _{a,b}$, respectively. Note 
that after the addition of the corresponding polynomial corrections, the 
bounds~(\ref{blowupfex}) and (\ref{grofainf}) are entirely analogous to the
bounds~(\ref{PWbound}) and (\ref{GSbound})---the operators $U_{\pm}$ are 
after all Fourier-like transforms~\cite{JPA04}.

Let us now apply the above theorems to the functions $\varphi ^{\rm in}(x)$
obtained by the ``standard method.'' In order for Eq.~(\ref{Upmac1D0inlrqFT}) 
to make sense, $\varphi ^{\rm in}(x)$ must fall 
off faster than exponentials. If we choose $\varphi ^{\rm in}(x)$ to fall off 
like $\rme ^{-|x|^a/a}$, then the Gelfand-Shilov theorem tells us that 
$\widehat{\varphi}^{\rm in}(q)$ grows like $\rme ^{+|{\rm Im}(q)|^b/b}$. Even 
when we impose that $\varphi ^{\rm in}(x)$ is $C_0^{\infty}$, which is already 
a very strict requirement, the Paley-Wiener theorem says that 
$\widehat{\varphi}^{\rm in}(q)$ grows 
exponentially. This means, in particular, that the 
$\widehat{\varphi}^{\rm in}(q)$ 
do in general {\it not} tend to zero in the infinite arc of the $q$-plane, 
because if they did, the Paley-Wiener and the Gelfand-Shilov theorems would be 
wrong. Because of Eq.~(\ref{Upmac1D0in2}), $\widehat{\varphi}^{\rm in}(z)$
does in general not tend to zero as $|z|$ tends to infinity in the lower 
half-plane of the second sheet. Hence 
the space of $\widehat{\varphi}^{\rm in}$'s is not of Hardy class from below. 

The space of $\widehat{\varphi}^{+}$'s cannot be of Hardy class from below 
either, because if it were, then 
\begin{equation}
      \lim _{|z|\to \infty} \widehat{\varphi}^{+}(z) =0 \, ,
          \label{sskdks}
\end{equation}
where the limit is taken in the lower half plane of the second sheet. By 
Eq.~(\ref{asstrco1}), this implies that also the space of
$\widehat{\varphi}^{\rm in}$'s would be of Hardy class and comply with this 
limit, which we know is not possible due to the ``classic results.'' Thus, 
the ``standard method'' yields spaces of test functions that 
do {\it not} comply with the Hardy axiom. This is precisely what it was 
meant in~\cite{HARDY} by the assertion that TAQT is inconsistent with standard 
quantum mechanics.

To finish this section, we note that if we chose the test functions as
in~\cite{BOLLINI}, then we would be dealing with ultradistributions. In
Fourier space, the test functions for ultradistributions grow faster 
than any exponential as we follow the imaginary axis, see~\cite{BOLLINI}
and references therein. Thus, if the 
``standard method'' yielded spaces of Hardy functions, that property of 
ultradistributions would be false.

\section{Further remarks}
\label{sec:further remarks}

The authors of~\cite{C} claim that it is inaccurate to state that the
proponents of TAQT dispense with asymptotic completeness. This statement should
be compared with the first quote in section~6 of~\cite{HARDY}.

The authors of~\cite{C} also claim that TAQT obtains the resonant states
by solving the Schr\"odinger equation subject to purely outgoing boundary 
conditions. This claim should be compared with the second quote in 
section~6 of~\cite{HARDY}.

The authors of~\cite{C} also dispute the assertion of~\cite{HARDY} that TAQT
sometimes uses the whole real line as though it coincided with the scattering
spectrum of the Hamiltonian. A glance at, for example, the 
QAT~(\ref{ferFtvph+}) seems to support such assertion.

\section{Conclusions}
\label{sec:con}

In standard scattering theory, one assumes that the physics is
described by the Lippmann-Schwinger equation. When one solves such equation,
one finds that its solutions must be accommodated by a rigged
Hilbert space, and that its time evolution runs from $t=-\infty$ till
$t=+\infty$~\cite{LS1}. When one analytically continues the solutions of
the Lippmann-Schwinger equation, one finds that they must be accommodated
by {\it one} rigged Hilbert space, which also accommodates the resonant (Gamow)
states. The construction of such rigged Hilbert space is determined by 
standard distribution theory.

By contrast, TAQT assumes that the solutions of the Lippmann-Schwinger 
equations belong to {\it two} rigged Hilbert spaces of Hardy
class. In TAQT, one never explicitly solves the Lippmann-Schwinger
equation for specific potentials in the position representation. Instead, one
assumes that its solutions satisfy the Hardy axiom. Unlike
in standard scattering theory, in TAQT the time evolution of the solutions 
of the Lippmann-Schwinger equation does not run from $t=-\infty$ till 
$t=+\infty$. 

By comparing the properties of the actual solutions of the Lippmann-Schwinger
equation with the Hardy axiom, we have seen that such actual solutions
would comply with the Hardy axiom only if classic results of 
Paley and Wiener, of Gelfand and Shilov, and of the theory of 
ultradistributions were wrong. We have (again) stressed the fact that the
Quantum Arrow of Time, which is the justification for using the rigged 
Hilbert spaces of Hardy class, has little to do with the time evolution of 
a quantum system. We have stressed that using the method of TAQT to 
introduce rigged Hilbert spaces, we could accommodate the Gamow states in
a landscape of arbitrary rigged Hilbert spaces, see also~\cite{RELBO}.

Our claim of inconsistency should not be taken as a claim that TAQT is 
mathematically inconsistent or that TAQT doesn't have a beautiful mathematical 
structure. What the present author claims 
is that TAQT is not applicable in quantum mechanics and is in fact a different 
theory.

To finish, we would like to mention that the ``classic theorems'' are not
in conflict with using Hardy functions in quantum mechanics. They are in
conflict only with the Hardy axiom. Thus, our results do not apply to other 
works that use Hardy functions in a different way~\cite{YOSHI}.

\ackn

This research was supported by MEC and DOE.

\end{document}